\documentstyle[12pt,epsfig]{article}
\newcommand\be{\begin{equation}}
\newcommand\ee{\end{equation}}
\newcommand\bea{\begin{eqnarray}}
\newcommand\eea{\end{eqnarray}}
\begin{document}

\title{Hamiltonian Flow Equations for the Lipkin Model}

\author{H.J. Pirner\thanks{
Supported by the Bundesministerium
f\"ur Bildung, Wissenschaft, Forschung und Technologie (BMBF) and GSI 
Darmstadt}\\
Institut f\"ur Theoretische Physik, Universit\"at Heidelberg\\
Heidelberg, Germany\\
and \\
B. Friman\\
GSI, Planckstr. 1, Darmstadt, Germany\\
and Institut f\"ur Kernphysik, TU Darmstadt, Darmstadt, Germany}


\maketitle

\begin{abstract} 
\setlength{\baselineskip}{18pt}
\noindent We derive Hamiltonian flow equations giving the
evolution of the Lipkin Hamiltonian to a diagonal form using
continuous unitary transformations.  To close the system of flow
equations, we present two different schemes. First we
linearize an operator with three pairs of creation and destruction
operators by reducing it to the $z$ component of the quasi spin. We
obtain the well known RPA-result in the limit of large 
particle number.  In the second scheme we introduce a new
operator which improves the resulting spectrum considerably 
especially for few particles.

\end{abstract}

\newpage

Recently, a new method \cite{We,Le} has been developed in condensed
matter physics to find approximate solutions in many-body theory. This
method is based on flow equations for the evolution of generalized
Hamiltonian couplings with a fictitious parameter $\ell$ running from
zero to infinity. The evolution parameter $\ell$ characterizes the
interval of energies $|E_i-E_j|<1/\sqrt{\ell}$, for which the
Hamiltonian has sizeable non-diagonal matrix elements. In the limit
$\ell\to\infty$ the Hamiltonian is diagonalized. The Hamiltonian flow
equations have been successfully applied to the problems of
electron-phonon coupling \cite{Le}, to BCS theory \cite{Mi} and
dissipative systems. It is natural also to try out this method on the
Lipkin model \cite{Li}, which is a standard test case for
approximation methods in nuclear physics. The Lipkin model is exactly
solvable, since its Hamiltonian is composed of bilinear products of
creation and destruction operators which have a quasi-spin
representation. Recently, new interest in the Lipkin model has
developed in the context of finite temperatures \cite{Ts} and as
a test of self-consistent RPA-type approximations \cite{Du}. 
The Hamiltonian flow equations for the Lipkin model can be
closed if one linearizes higher products of quasi-spin
operators into a product of expectation values and a quasi-spin
operator.  When the number of particles $N$ is very large we naturally
reproduce the RPA results for $N\to\infty$.  Extra care is necessary
to achieve satisfactory results for mesoscopic problems with $N=(8,10
..20)$ particles. We will give an example how to improve the simple
linear scheme.

In the Lipkin model $N$ particles can distribute themselves on two
levels, which are both $N$-fold degenerate. The splitting of the
levels is $\xi_0$ and the interaction $V_0$ mixes states where two
particles simultaneously move from the lower level to the higher level
or vice versa:  
\be\label{1} {\cal
H}=\frac{1}{2}\xi_0\sum_{\sigma,p}\sigma a^+_{p\sigma} a_{p\sigma}
+\frac{1}{2} V_0\sum_{pp',\sigma} a^+_{p\sigma} a^+_{p'\sigma} a_{p'-
\sigma} a_{p-\sigma}
\ee 
In the Hartree-Fock ground state all particles
are in the lower level, where they have the quasi-spin $\sigma=-1$.
Therefore the total quasi-spin $J_z$, \be\label{2}
J_z=\frac{1}{2}\sum_{p,\sigma}\sigma a^+_{p\sigma} a_{p\sigma},\ee has
the value $\langle J_z\rangle=-\frac{1}{2} N$.

The interaction term can also be expressed in terms of
quasi-spin operators. The Hamiltonian ${\cal H}$ then has the form
\be\label{6} 
{\cal H}=\xi_0 J_z+V_0(J^2_++J^2_-),
\ee
where
\be \label{3}
J_+=\frac{1}{\sqrt2}\sum_p a^+_{p,+1} a_{p,-1},\
J_-=\frac{1}{\sqrt2}\sum_p a^+_{p-1} a_{p+1},
\ee 
and $J_z$ form an angular momentum algebra 
\bea  \label{4}
[J_z,J_\pm]&=&\pm J_\pm;\nonumber\\{}
[J_+,J_-]&=&J_z.
\eea 
Since $[\vec{J}^{\, 2},{\cal H}]=0$, the interaction mixes only states in the
quasi-spin multiplet $J=\frac{1}{2}N$.
In the following we will consider a limited range of couplings such
that the Hartree-Fock ground state remains stable,
i.e. $\varepsilon=\frac{NV_0}{\xi_0}<1$, otherwise one would have to transform
to a deformed basis \cite{Du}.

The concepts of unitary transformations to diagonalize the Hamiltonian
is well established. Negative energy states can be eliminated from the Dirac
equation by unitary transformations \cite{Fo,bp}. The $e^S$ method 
\cite{Bi} with a suitable ansatz for the operator $S$
has been extremely successful to derive the hierarchy of two-, three-
and more-body correlations in many-particle systems, especially
nuclei.  The new idea in the Hamiltonian flow equations is to make the
unitary transformation infinitesimal, so that it can adjust itself
best to diagonalize the Hamiltonian at each stage of the flow.

In the field theory of critical phenomena it is advantageous 
to gradually integrate out momentum shells 
instead of eliminating the whole high momentum region in one step.
In Hamiltonian theory the energy representation is
the most useful representation for unitary transformations to
eliminate non-diagonal matrix elements of shorter and shorter
range. The resulting couplings in ${\cal H}$ and therefore the
Hamiltonian itself will then carry an index $\ell$, labeling the
stage of diagonalization. With $\tilde\eta$ anti-hermitian the unitary
transformed ${\cal H}$ has the following form \be\label{7} {\cal
H}'=e^{\tilde\eta} {\cal H} e^{-\tilde\eta}.\ee In the case of an
infinitesimal $\tilde\eta=\eta\Delta\ell$ we approximate the change
\be\label{8} {\cal H}(\ell+\Delta\ell)-{\cal
H}(\ell)=[\eta(\ell),{\cal H}(\ell)]\Delta\ell\ee and obtain the
Hamiltonian flow equation: 
\be\label{9} 
\frac{d{\cal
H}(\ell)}{d\ell}=[\eta(\ell),{\cal H}(\ell)].
\ee 
In general one chooses $\eta$
proportional to the commutator of the diagonal part of the Hamiltonian
${\cal H}_D$ and $\cal H$
in order to decrease the magnitude of the off diagonal
matrix elements during evolution: \be\label{10} \eta(\ell)=[{\cal
H}_D(\ell),{\cal H}(\ell)].\ee It is easy to follow the non-diagonal
matrix elements of ${\cal H}$ in the course of evolution with $\ell$:
\be\label{11} 
\frac{d{\cal H}_{ij}}{d\ell}=\sum_k(\eta_{ik}{\cal
H}_{kj}- {\cal H}_{ik}\eta_{kj}) \ee and \be\label{12}
\eta_{ik}=(E_i-E_k){\cal H}_{ik}\ee gives \be\label{13} \frac{d{\cal
H}_{ij}}{d\ell}=-(E_i-E _j)^2{\cal H}_{ij}+{\cal O} \left(({\cal
H}_{ij})^2\right).  
\ee 
The leading term in eq.~(\ref{13}) arises by
using the diagonal part of the Hamiltonian in eq.~(\ref{11}). The
off-diagonal part of the Hamiltonian yields sub-leading terms. The
resulting equation suggests a damping of the non-diagonal elements
which decrease with increasing $\ell$.  \\

Let us first take the original Hamiltonian with $\ell$-dependent
couplings and study its evolution with the Hamiltonian flow
equations. We improve the scheme in the second half of the
paper by adding a term which starts out as zero for $\ell=0$
but grows during the evolution. The generalized Hamiltonian ${\cal
H}(\ell)$, together with the initial conditions, has the following form:
\bea\label{14}
{\cal H}(\ell)&=&\xi(\ell)J_z+V(\ell)(J^2_++J^2_-),\nonumber\\
\xi(0)&=&\xi_0,\nonumber\\
V(0)&=&V_0.
\eea
Defining   $\eta$ as
$
\eta=[{\cal H}_D,{\cal H}]$
with
$
{\cal H}_D=\xi(\ell)J_z
$
we obtain:
\be
\eta= 2 \xi V (J^2_+-J^2_-).
\ee
By inserting this $\eta$ into the flow equation we find:
\bea\label{20}
\frac{d{\cal H}}{d\ell}&=&[\eta, {\cal H}]\nonumber\\
                       &=&4 \xi V^2( 2 J_z(\vec J^2-J_z^2)-J_z)
-4\xi^2 V(J^2_++J^2_-). 
\eea
The right hand side of this equation contains an hermitian operator
containing three quasi spins $\propto J_z^3$, i.e. a term different
from the original evolving Hamiltonian (\ref{14}). 
This new term is approximated
by linearizing around the Hartree-Fock operator expectation values:
\be
A B  = (A- \langle A \rangle)(B- \langle B\rangle)-
\langle A \rangle\langle B\rangle
+\langle A \rangle B +
A \langle B\rangle
\ee 
which yields:
\be
J_z^3 \rightarrow 3 \langle J_z\rangle ^2  J_z.
\ee
A
linearization of $J^3_z$ is certainly meaningful when the fluctuations
of $J_z$ around its expectation value are small, i. e. for large quasi
spins.  This is the case for the lowest lying levels, if the particle
number is large.

Differentiation of the ansatz for ${\cal H}(\ell)$ 
in eq. (\ref{14}) gives the left side of the evolution equation
\be\label{21}
\frac{d{\cal H}}{d\ell}=\frac{d\xi}{d\ell} J_z+\frac{dV}{d\ell}(J^2_++J^2_-)
\ee
which we compare with the approximate result for the commutator 
from eq. (\ref{20}).

We see that the  ansatz for ${\cal H}(\ell)$ in eq. (\ref{14})  together with  
the
approximation for the cubic operator product indeed leads to a closed system 
of
first order differential equations  for the coefficients in front of
the quasi-spin operators.
\bea\label{24}
\frac{d\xi}{d\ell}&=&-4 \xi V^2[ N(N-1)+ 1],\nonumber\\
\frac{dV}{d\ell}&=&-4 \xi^2 V.
\eea
The second equation implies that 
the magnitude of the
off-diagonal matrix element $V$ decreases in the course of the
evolution, independently of its sign.
Combining the two equations one readily obtains an invariant of the
evolution:
\be
\xi^2(\ell)-[N(N-1)+ 1]V^2(\ell) = \xi^2_0-[N(N-1)+ 1]V^2_0.
\ee 
Using $V(\infty)=0$
one finds 
the limiting value of $\xi$ and the splitting $\Delta$ between 
the ground state and first excited state :
\bea
\xi(\infty)&=&\xi_0 \sqrt{1-\bar \varepsilon^2},\nonumber\\
\bar \varepsilon&=&\sqrt{N(N-1) +1 }\frac{V_0}{\xi_0};\nonumber\\
\Delta&=&\langle 1|{\cal H}(\infty)|1\rangle-\langle 0|{\cal
H}(\infty)|0\rangle\nonumber\\
&=&\xi(\infty).
\eea 

In the large $N$-limit with finite interaction strength 
the
splitting $\Delta$ between the ground state and first excited state has the 
form of the RPA approximation \cite{Li}
\bea\label{28}
\Delta_{RPA}&=&\xi_0 \sqrt{1-\varepsilon^2},\label{26}\nonumber\\
\varepsilon&=&N\frac{V_0}{\xi_0}.
\eea

In nuclear physics one studies systems with intermediate particle
number. Neither few nor many body methods are directly applicable.
In nuclei the relevant number of particles is given by the number of  
particles
outside of closed shells, which is $N \leq 2J+1$, where $J$ is the
angular momentum of the valence shell.   
We want to obtain a reliable result also for small values of $N$. 
To this end we propose to add a new operator to the original
Hamiltonian, which improves the accuracy of the evolution equations.
The new term $\propto J_z^3$ has been chosen to have the same
symmetry
as the original Hamiltonian under reflections. From the
previous flow equations such an improvement is suggested, since
a term $\propto J_z^3$ was generated on the 
right hand side of the previous evolution equation (\ref{20}).
Adding new operators generated from the original 
Hamiltonian by the commutators is an established way to improve the method. 
Obviously symmetry provides an important guideline for the choice of
new operators to be included in the Hamiltonian ${\cal H}(\ell)$.
Usually the commutators generate  
additional terms which must be eliminated, since one can
obtain a closed set of differential equations only if every 
operator on the right side of the evolution equation has a counterpart
on the left-hand side of eq. (\ref{9}).  We choose:
\be\label{14a}
{\cal H}(\ell)=\xi(\ell)J_z+V(\ell)(J^2_++J^2_-)+
\gamma(\ell) J_z^3
\ee
with the initial conditions
\be\label{15}
\xi(0)=\xi_0;\ V(0)=V_0;\ \gamma(0)=0.\ee

Proceeding along similar lines as before we define  $\eta$ as
\be\label{17}
\eta=[\xi J_z,{\cal H}]\ee
slightly deviating from the commutator of $[{\cal H}_D,{\cal H}]$.
Otherwise terms with higher 
powers of quasi spins would be produced. 
The explicit calculation of $\eta$ leads to the same anti-hermitian operator 
product as before.
\be
\eta= 2 \xi V (J^2_+-J^2_-).
\ee
Inserting this $\eta$ into the flow equation we get

\bea
\frac{d{\cal H}}{d\ell}=4 \xi V^2( 2 J_z(\vec J^2-J_z^2)-J_z)
-4\xi^2 V(J^2_++J^2_-)+2 \xi V \gamma
[(J^2_+-J^2_-),J_z^3].
\eea
We see  a new operator product appearing from the
commutator of $\eta$ with the $\gamma$ term in ${\cal H}$.
Using the reduction scheme for products of operators described above,
we find
\be
[(J^2_+-J^2_-),J_z^3] \rightarrow  -6 \langle J_z \rangle ^2
(J^2_++J^2_-)
\ee
This way we regain a closed system 
of
first order evolution 
equations  for the coefficients in front of the quasi-spin
operators
\bea\label{24a}
\frac{d\xi}{d\ell}&=&4 \xi V^2[ 2 J(J+1) -1], \nonumber\\
\frac{dV}{d\ell}&=&-4 \xi^2 V -12 \xi V \gamma \langle J_z 
\rangle ^2,\nonumber\\
\frac{d\gamma}{d\ell}&=&-8 \xi V ^2.
\eea

We introduce the abbreviations 
\bea
r&=&2 \vec J^2 -1 = 2J(J+1) -1,\nonumber\\
s&=&\frac{3}{4}\langle J_z \rangle ^2, \nonumber\\
t&=&\frac{8s}{r}.
\eea
Combining the first and third evolution equation we find the invariant
\be
\frac {\xi}{r} +\frac{\gamma}{2}=\frac {\xi_0}{r},
\ee
where we employ the initial conditions on the right hand side.
By using this relation we eliminate $\gamma$ as independent variable.
A second more complicated invariant arises from all three
equations and has the form
\be
\xi^2+r V^2-t(\xi_0-\xi)^2=\xi_0^2+rV_0^2.
\ee
\begin{figure}[tbh]
\setlength{\unitlength}{1mm} 
\begin{picture}(150,80)
\put(25,0){\epsfig{file=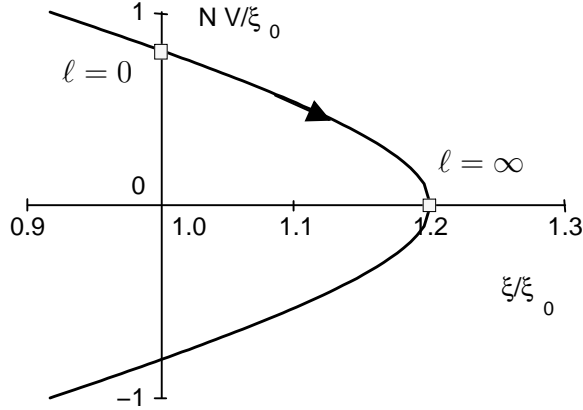,height=80mm}}
\put(45,55){$\ell=0$}
\put(95,43){$\ell=\infty$}
\end{picture}
\caption{Flow in the $\xi -V$ plane as defined by eqs.~(29) and (32)}
\end{figure}
The allowed values of $\xi$ and $V$ lie on a hyperbola as shown in
fig. 1.
For the plot we choose $\varepsilon=0.8$ and $t =3$, which is 
approximately true for large $N$.
At the beginning of the evolution at 
$\ell=0$ we start on the hyperbola at $\xi=\xi_0$ and
$V=V_0$. Since the gradient ${d\xi}/{d\ell}>0$ the 
magnitude of $\xi$ is growing and $V$ decreases. Thus the point $V=0$,
where the right hand sides
of all three evolution equations vanish, is a stable fix point. 
The variables $\xi$
and $\gamma$ converge towards the fix point values  
\bea
\xi(\infty)&=&\frac{\xi_0}{t-1}\left[
t-\sqrt{1-3\left(\frac{t-1}{2t}\right)\varepsilon^2}\,\right],\\
\gamma(\infty)&=&\frac{2\xi_0}{r(t-1)}\left[-1+
\sqrt{1-3\left(\frac{t-1}{2t}\right)\varepsilon^2}\,\right].
\eea
The above expressions depend on the particle number $N$ 
through the
constants $r,s,t$ in a rather complicated way. 
Due to the presence of the $J_z^3$ term in the Hamiltonian the
splitting $\Delta$ between the ground state and first excited state has the 
modified  form:
\bea\label{28a}
\Delta&=&\langle 1|{\cal H}(\infty)|1\rangle-\langle 0|{\cal
H}(\infty)|0\rangle\nonumber\\
&=&\xi(\infty) + \gamma(\infty) (\frac{3N^2}{4}-\frac{3N}{2} +1 )
\eea

\begin{figure}[tbh]
\setlength{\unitlength}{1mm} 
\begin{picture}(150,185)
\put(25,130){\epsfig{file=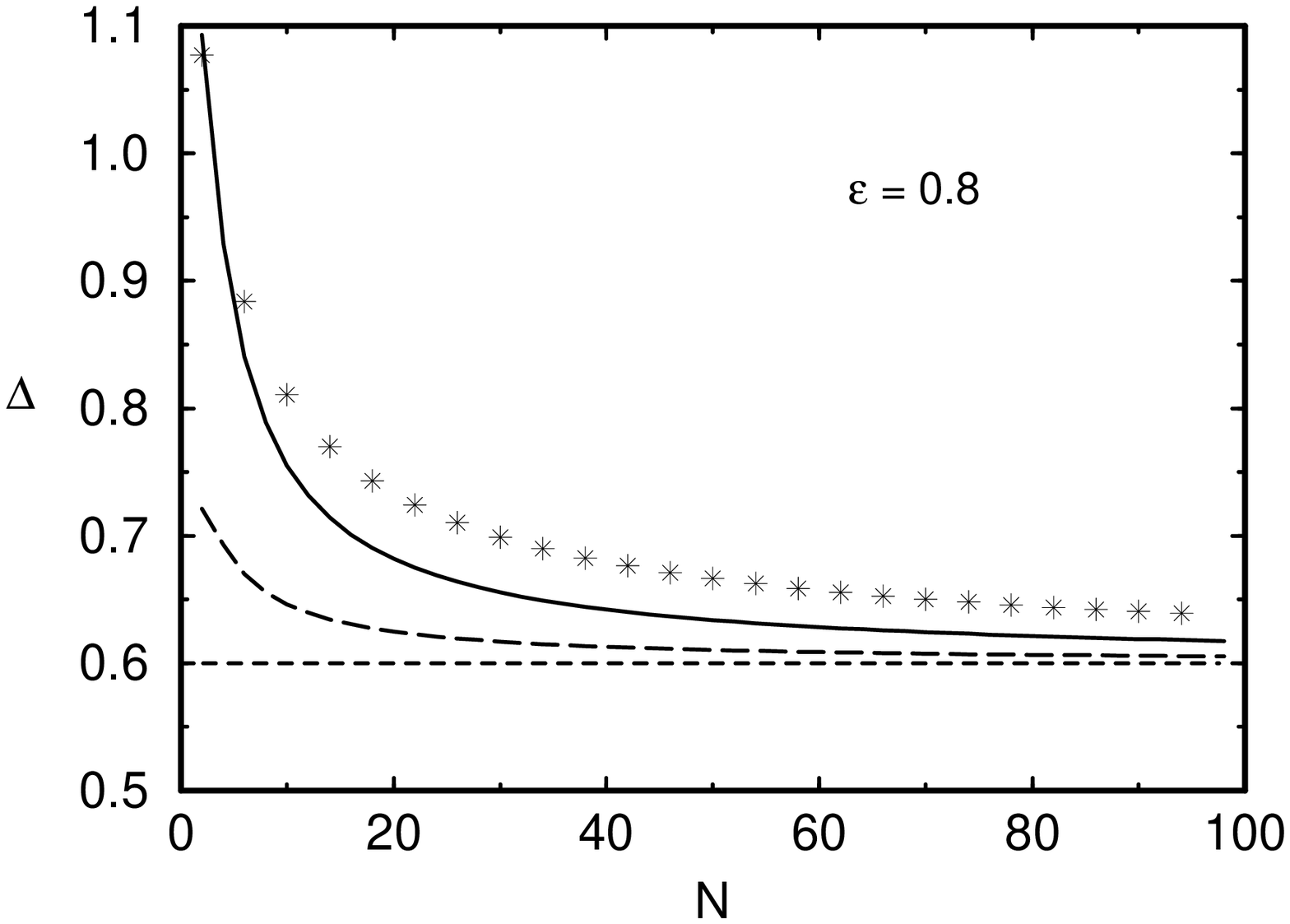,height=80mm}}
\put(25,65){\epsfig{file=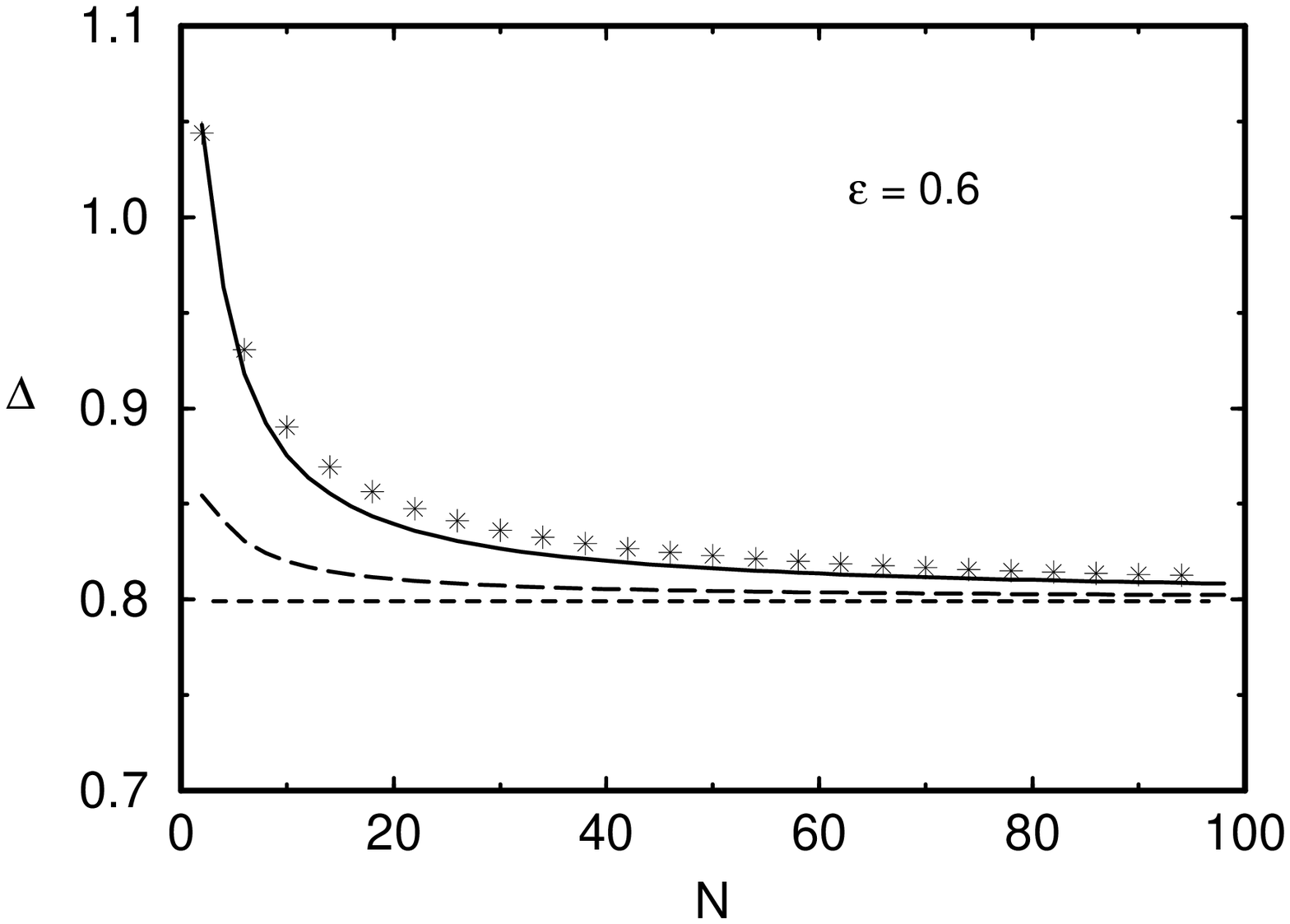,height=80mm}}
\put(25,0){\epsfig{file=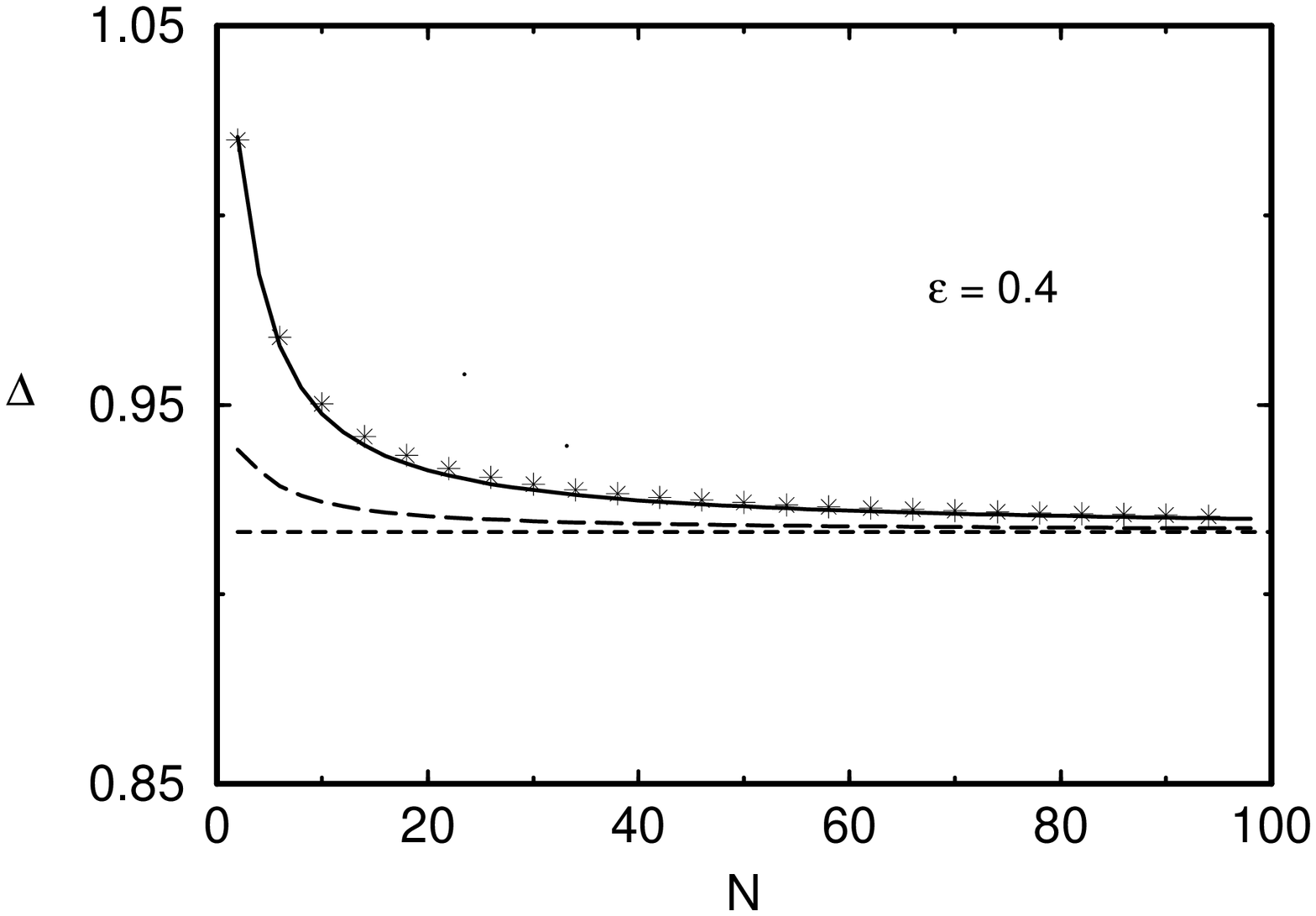,height=80mm}}
\end{picture}
\caption{Energy gap $\Delta$ between ground state and first excited
state for various strengths $\varepsilon= 0.4, 0.6, 0.8$. Stars denote
the exact result, solid line shows the improved result from the evolution
equations, while the long-dashed line gives the result of the
linearized scheme. The short-dashed line is the RPA result.}
\end{figure}
In fig.2 we show the energy gap $\Delta$ for three different interaction
strengths $\varepsilon= 0.4, 0.6, 0.8$. The curves with stars
are the result of an exact diagonalization of the Hamiltonian.
The full drawn curves give the improved result from the evolution
equations (\ref{24a}). 
For comparison we also show the result of the evolution equations
(\ref{24}) with long dashes and the RPA result (\ref{28}) with short dashes.
One sees that the method
is very accurate for weak interactions, where perturbation theory also
does a rather good job.
In the case of strong coupling there are still some deviations for intermediate
particle numbers.
The larger the coupling the closer one is to
the transition to a deformed ground state, therefore
also the fluctuations become stronger 
and our expansion of the new operators in 
powers of fluctuating operators is not so good anymore.
All three calculations approach the RPA result
for large particle numbers shown with short dashes. 
In the limit of large $N$ one can also 
demonstrate analytically
that the improved method converges to the RPA result.

Applying the evolution equations to the Lipkin model 
one encounters the problem that ever new operators are generated 
in the course of setting up the evolution equations. This is
a general feature of spin models. Similar features occur in 
the spin boson problem. The original Lipkin Hamiltonian has a rather 
simple tridiagonal structure, which is lost during the evolution for finite
$\ell$, if one does not enforce it by suitable approximations. 
Recently Mielke \cite {mi2} has 
constructed an $\eta$-operator which preserves the
simple structure of the Lipkin Hamiltonian without approximations, but still 
reduces the magnitude of
the off-diagonal matrix elements. This method is also applicable beyond
the stability of the Hartree-Fock state $(\varepsilon>1)$. 

In conclusion, we have set up Hamiltonian
flow equations for the Lipkin model.
By infinitesimal unitary transformation we have diagonalized
the Hamiltonian approximately making reasonable truncations for
the new operators. We have presented an improved scheme 
which successfully interpolates between perturbation theory 
for small $N$ and RPA for large $N$. The method of evolution equations 
has been shown to be a promising new
tool also for nuclear physics. It opens up new perspectives for
approximations in many body problems which still have to be
explored in greater detail. Thus a systematic approach 
to select the relevant set of operators for the evolution equations
can be established.

\section*{Acknowledgements}
We thank F. Wegner and A. Mielke for their help and expertise 
in flow equations communicated to us.

\end{document}